\documentclass{article}

\PassOptionsToPackage{numbers}{natbib}
\usepackage[final]{nips_2016}

\usepackage[utf8]{inputenc} 
\usepackage[T1]{fontenc}    
\usepackage{hyperref}       
\usepackage{url}            
\usepackage{booktabs}       
\usepackage{amsfonts}       
\usepackage{nicefrac}       
\usepackage{microtype}      
\usepackage{graphicx}
\usepackage{amsmath, amssymb}

\title{Hierarchical Detail Enhancing Mesh-Based Shape Generation with 3D Generative Adversarial Network}

\author{
  Chiyu `Max' Jiang \\
  University of California\\
  Berkeley, CA 94720 \\
  \texttt{chiyu.jiang@berkeley.edu} \\
  \And
  Philip Marcus \\
  University of California\\
  Berkeley, CA 94720 \\
  \texttt{pmarcus@me.berkeley.edu} \\
}
\begin{document}

\maketitle
 
\begin{abstract}
Automatic mesh-based shape generation is of great interest across a wide range of disciplines, from industrial design to gaming, computer graphics and various other forms of digital art. While most traditional methods focus on primitive based model generation, advances in deep learning made it possible to learn 3-dimensional geometric shape representations in an end-to-end manner. However, most current deep learning based frameworks focus on the representation and generation of voxel and point-cloud based shapes, making it not directly applicable to design and graphics communities. This study addresses the needs for automatic generation of mesh-based geometries, and propose a novel framework that utilizes signed distance function representation that generates detail preserving three-dimensional surface mesh by a deep learning based approach.
\end{abstract}

\section{Introduction}
\label{sec:intro}
Automatic generation of three-dimensional shapes is of dear interest to disciplines such as computer graphics (CG), where a vast amount of new shapes is required to populate a virtual world. 3D model generation is of greater renewed interest thanks to the development of virtual and augmented reality technology, where real-time shape analysis and synthesis by computer vision techniques is required.

The past decade has seen the development of a variety of techniques for addressing this need, mostly originating from researchers in the computer graphics community. However, most of the work in this direction are related to so-called assembly based 3D modeling, which creates new shapes based on mix-and-matching different parts from a database of 3D models. Besides shape generation, the computer graphics research community has developed a suite of tools for post-processing and rendering of 3D mesh, making it possible for artists in the gaming and movie industry to create photorealistic 3D models based on polygonal mesh data-structures. Though such methods manage to create new out-of-sample models that are realistic looking with high levels of details, they suffer from two major drawbacks, namely their inability to create conceptually novel shapes, and the lack of a high-level global descriptor for generalized shapes. 

Recent advances in deep-learning based computer vision (CV) research lead to a different approach towards shape generation. New work along this line of research utilize deep learning frameworks, most predominantly convolutional neural networks (CNN) to encode 3D shape information and also to synthesize novel shapes from latent shape vectors that encodes a continuous space of 3D shapes. These methods appeal to researchers for various reasons. First, these methods are capable of producing novel shapes beyond simple recombination of parts from a database. Second, these methods are valuable not only for the synthesis of new shapes, but also for the analysis of shapes, since it provides an encoding scheme that encodes 3D shapes to a latent space containing high-level feature abstractions. Moreover, these learning-based methods can be trained in an end-to-end fashion, eliminating the need for complicated heuristics-based methods. However, most published results along this line of research utilize volumetric occupancy voxel grids for the representation of input as well as generated shapes. Though voxel grids offer a lean, memory-efficient and convolution-ready means of representing 3D shapes, the resulting shapes lack general visual appeal. Moreover, it is impossible to post-process voxel based geometries with current mesh-oriented algorithms. Other works alternatively use a point-cloud based representation for better integration with lidar sensors for autonomous driving related applications, but the visual qualities of its results are even further from applications in CG and design.

In light of current progress in CG and CV, we propose a novel deep learning based hierarchical scheme for the generation of visually appealing 3D mesh-based shapes with enhanced details using CNN that works with signed distance function representation of shapes. Instead of generating voxel occupancy level shape information as in \cite{3dgan} \cite{tatarchenko2017octree}, we propose to instead generate signed distance function fields on gridded domain using a similar architecture consisting of up-convolutional layers, for easy reconstruction of polygon mesh surface, which is of much higher quality compared to voxel-based models. We further propose a hierarchical approach towards shape generation by separating the frequencies in the resulting fields, and using two different generators to generate the different frequency portions of the image. Resulting shapes from our recent study are smooth and visually appealing, featuring high-resolution details. We make the argument that in the case of GAN assisted shape generation, higher grid resolution does not necessarily amount to higher quality.

In the following sections, in section 2, we will discuss related work in the literature. in section 3, we will discuss various details related to geometry representation and processing. In section 4, we will discuss our model architecture and training details. In section 5, we present results synthesized by our model. In section 6, we wrap up this study with conclusions and future directions. 

\section{Related Work}
\subsection{Assembly Based Shape Generation}
Assembly based shape generation techniques create new shapes by segmenting geometries in a database and recombine different parts to form new shapes. Such algorithms typically use a pipeline of mesh-based algorithms for shape analysis and synthesis. Shape analysis consists of shape segmentation, contact analysis and hierachy analysis of different parts. Shape synthesis typically consists of shape matching and contact enforcement. \citet{jain2012exploring} used such analysis and synthesis pipeline to generate new shapes by linear interpolation of baseline shapes. \citet{kalogerakis2012probabilistic} furthermore used a learning based approach for automated combination of parts in a dataset, leading to creation of more realistic and coherent shapes. However, such generations schemes are inherently unable to generate conceptually novel shapes, or offer a more high-level representation for these shapes

\subsection{Convolution Based Shape Generation}
The use of 3D up-convolutional schemes for shape generation was, to our best knowledge, first proposed by \citet{3dgan} to generate voxelized volumetric shapes. Such schemes utilize the up-convolutional operators in deep learning, also sometimes known as transposed convolution or deconvolution, for generating a function field on a gridded domain. \citet{3dgan} combined the use of Generative Adversarial Networks \cite{goodfellow2014generative} with 3D up-convolution based decoders for the unsupervised learning of 3D shape representation as well as the generation of new shapes. A stream of followup studies have looked into the use of such architectures for related tasks such as shape completion \cite{dai2016shape}, image-to-shape translation \cite{hane2017hierarchical} and interective 3D modeling \cite{liu2017interactive}. A more recent point of focus is in the use of octree based up-convolution operations, as such memory efficient implementation allows a push for higher spatial resolution of up to $256^3$ \cite{tatarchenko2017octree} \cite{hane2017hierarchical}.

\subsection{Detail Enhancement by Nearest Neighbor}
Some work in the literature have used nearest neighbor search for shape detail enhancement. \citet{dai2016shape} performed detail enhancement using a patch-based $k$-nearest neighbor query to refine the low-resolution input in $\mathbb{R}^{32\times 32\times 32}$ to $\mathbb{R}^{128\times 128\times 128}$. However such nonparametric models suffer from slow query during test time and requires the availability of the entire shape database.

\section{Geometry Processing}
\label{sec:geom}
In this section, we present a robust geometry processing pipeline that is used to convert large amount of mesh based shape data to a learnable format (3D Signed Distance Function field). This section is dedicated to providing a brief introduction to the properties of the signed distance function and an outline of the pipeline we used for performing data conversion.

\subsection{Signed Distance Function}
Signed distance functions (SDF) have been widely used in the computer vision community for applications such as rendering and segmentation. A mathematical definition of the signed distance function is given as follows:
\begin{gather}
f(x)=
\begin{cases}
d(x, \partial \Omega) \qquad 
\text{if}x\in\Omega \\ 
-d(x, \partial \Omega) ~\quad\text{if}x\in\Omega^{c}
\end{cases}
\end{gather}
where $\partial \Omega$ denotes the boundary of $\Omega$. For any $x\in X$,
\begin{equation}
d(x, \partial \Omega):=\inf_{y\in\partial \Omega}d(x, y)
\end{equation}
where $\inf$ denotes the infimum. In Euclidean space, the signed distance function has some desirable properties, such as for piecewise smooth boundary, the signed distance function is differentiable almost everywhere, and its gradient satisfies the Eikonal Equation:
\begin{equation}\label{eqn:eikonal}
|\nabla f| = 1
\end{equation}
\subsection{Processing Algorithm}
This study mainly involves the use the ShapeNet dataset \cite{shapenet2015}. The shape data in the dataset is given as triangular mesh, hence a geometry processing pipeline is needed for the conversion of triangular mesh to gridded signed distance field.

First, we center and normalize the imported triangular mesh and set up a unit size 3D spatial grid around the geometry. Then we use the data structure AABB (Axis-aligned Bounding Box) tree for efficient point-to-mesh distance query. After calculating the point to mesh distance everywhere in the domain, we compute the winding number \cite{jacobson2013robust} for each point in the domain to determine the sign. The algorithm is robust in that it is functional even in the case of a non watertight mesh.

In this study, we leveraged the computational geometry code infrastructure provided in the open-source library libigl \cite{libigl} to facilitate this process.
\section{Model}
\label{sec:model}
In this section, we present an overview of the hierarchical models we adopted for shape generation. The architecture consists of two separate networks, mainly the low-frequency generator (LFG) and the high-frequency generator (HFG). We also describe the training procedures involved.

\begin{figure}
\centering
\includegraphics[width=\textwidth]{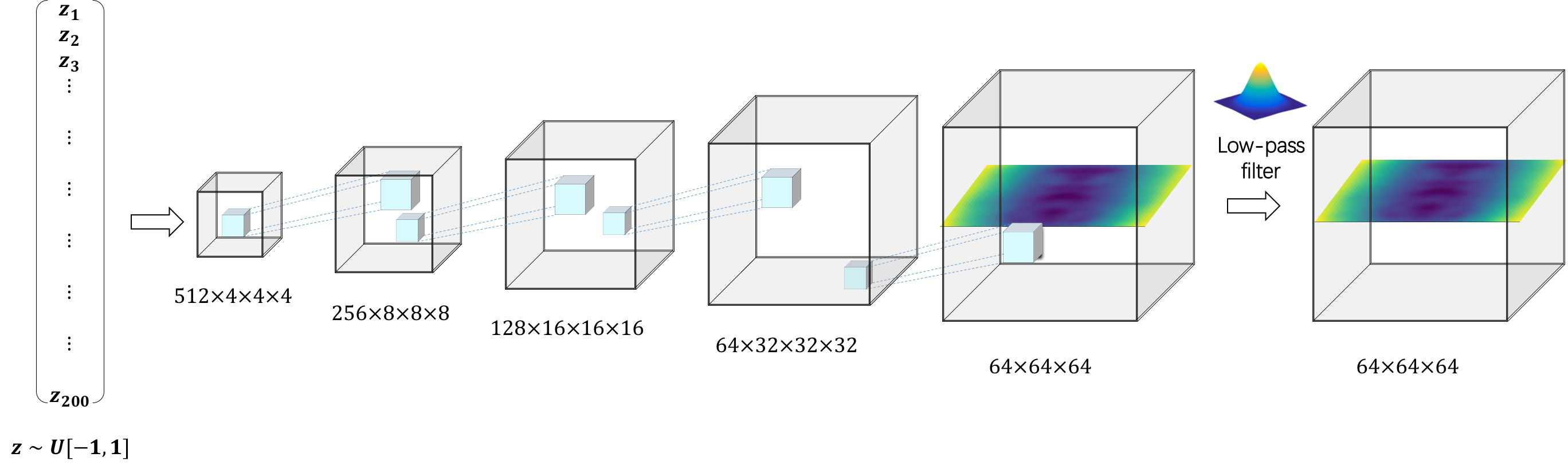}
\caption{Schematic for the Low-Frequency Generating network. The network takes in a random vector $z\sim U[-1, 1]$ and use up-convolution layers to project to $\mathbb{R}^{64\times 64\times 64}$. During test time, the results are passed through a low-pass filter to extract low-frequency components. The discriminator for LFG is a mirror of the above architecture, excluding the low-pass filter.}
\label{fig:lfg}
\end{figure}

\begin{figure}
\centering
\includegraphics[width=\linewidth]{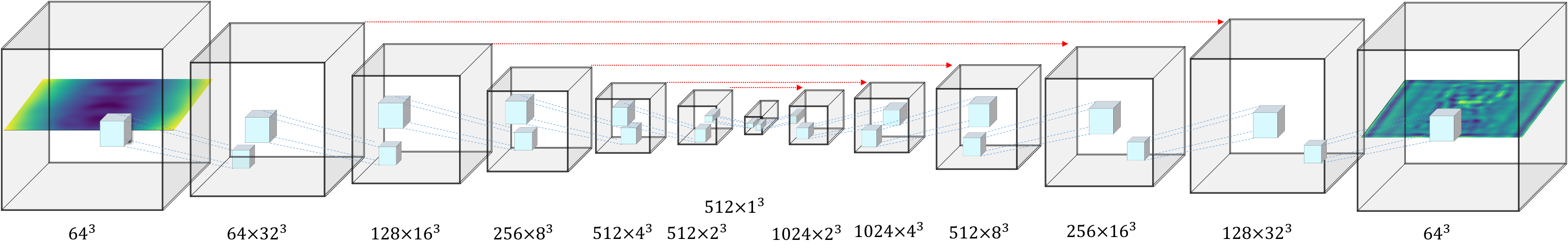}
\caption{High-Frequency Generator (HFG) Architecture. HFG is a conditional GAN that includes skip layers. The HFG takes in low-frequency inputs from the dataset and map it to the corresponding high-frequency image from the dataset.}
\label{fig:hfg}
\end{figure}

\subsection{Low Frequency Generator}
The architecture of the LFG is inspired by that of the two dimensional deep convolutional generative adversarial network (DC-GAN) \cite{radford2015unsupervised} and voxel based 3D-GAN \cite{3dgan}. Like the previous two examples, our network utilizes an up-convolutional neural network for the generation of signed distance field. Different from the previous examples, we perform low-pass filtering on the results to reduce noise in the high-frequency domain and use it as an input for HFG to generate high frequency details.

The overall loss can be written as a sum of the losses from the discriminator and the generator, given as:
\begin{equation}
L_{GAN}=\log D(x)+\log (1-D(G(z)))
\end{equation}
where $x$ is the value of the signed distance function in $64\times 64\times 64$ of a real sample in the training set, $z$ is a 200-dimensional vector that is sampled i.i.d. from a uniform distribution over $[-1, 1]$. $D(x)$ is the output from the discriminator, and $G(z)$ is the output from the generator in $64\times 64\times 64$.

The generator takes in the vector $z$ and projects the vector to a higher dimension using linear layer and reshapes it ($512\times 4\times 4\times 4$). It then passes through four up-convolution operations with kernel size $5\times 5\times 5$ and strides 2, with batch normalization and followed by ReLU layer after each convolutional layer. The result is then passed through the hyperbolic tangent (tanh) activation function to comply with the output range of -1 and 1. Though not a part of the training process, the generated results are post-processed by passing them through a low-pass gaussian filter and then fed to the High-Frequency Generator. A triangular-meshed surface can be further extracted from the SDF field using the marching cubes algorithm. An illustration of the Low-Frequency Generator can be found in Figure \ref{fig:lfg}.

The discriminator is an almost mirror image of the generator network, with the difference being that instead of using ReLu, it utilizes leaky ReLU with slope of 0.2. Leaky Relu is formulated as:
\begin{equation}\label{eqn:lrelu}
LReLU(x) = \max(x,\alpha x)
\end{equation}
where $\alpha < 0$ is the slope.

\subsection{High Frequency Generator}
High Frequency detailed are generated by the High Frequency Generator, which is a conditional GAN conditioned on low frequency inputs. The architecture of the HFG is inspired by that of the two dimensional pix2pix network \cite{isola2016image} that performs image-to-image translation. Similar to the goal of the pix2pix network, the objective of the HFG is to perform image to image translation from low frequency to high frequency on a per pixel basis. Hence, the network is signature by an encoding part, a decoding part, and skip layers between the encoder part and the decoder part to pass through per-pixel level information.A schematic for the generator of HFG is presented in Figure \ref{fig:hfg}. The discriminator architecture for HFG is similar to that of the LFG, except that the image is penalized by a patch-level penalty, meaning that the discriminator is tasked to distinguish between real and fake image patches rather than entire images. This better preserves local high-frequency structures.

The loss for the HFG network consists of generator loss, discriminator loss, and a L1 loss which be written as follows:
\begin{equation}
L_{GAN}=\log D(x_{lf}, x_{hf})+\log (1-D(x_{lf}, G(x_{lf})))+||x_{hf}-G(x_{lf})||_{1}
\end{equation}
where $x_{lf}$ and $x_{hf}$ are the low frequency and high frequency parts of the same image from the training set. $D(x_{lf}, x_{hf})$ is the output of the discriminator when presented with a low-frequency and high-frequency pair. The encoder section uses a leaky Relu as activation function after each convolution operation, and the decoder section uses a Relu as activation before each up-convolution.

\begin{figure}[!t]
\centering
\includegraphics[width=\textwidth]{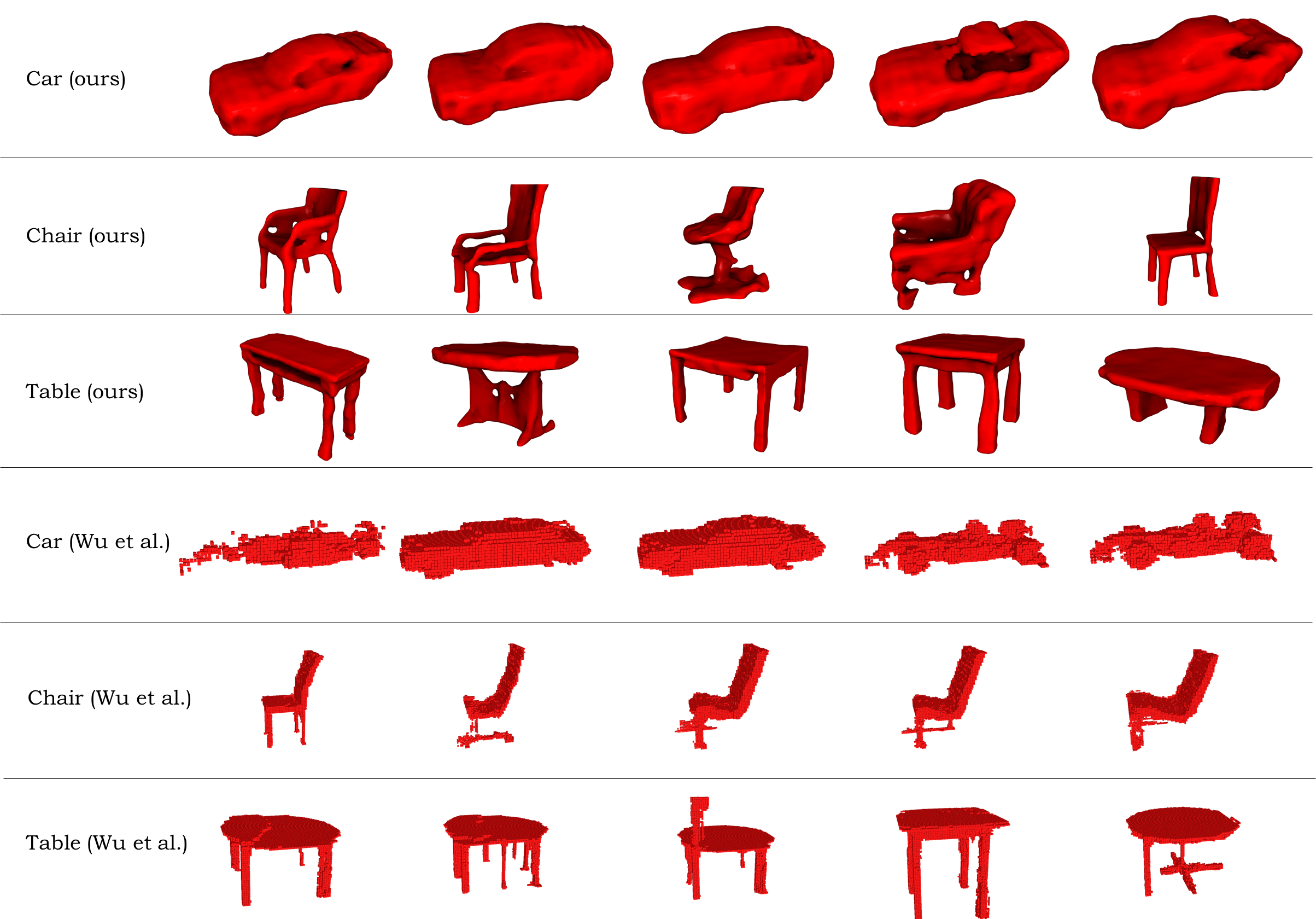}
\caption{A sample of the shapes generated by our model using 200 dimensional random vectors sampled uniformly from $U[-1, 1]$. The top three rows show our generated samples and bottom three rows show samples generated by \citet{3dgan} from the corresponding classes.}
\label{fig:collage}
\end{figure}

\subsection{Training Details}
The LFG and HFG are trained separately. Using different data partitions. The LFG is trained using the original Signed Distance Function in $\mathbb{R}^{64\times 64\times 64}$ gridded domain. The HFG is trained using the same dataset, but after splitting it into low-frequency and high-frequency pairs, since it is a conditional GAN for learning low-frequency to high-frequency mapping. We use the Adam optimizer \cite{kinga2015method} for the training of both the generator and the discriminator. In order to stabilize the training process and prevent the discriminator from overpowering the generator, we use a combination of two approaches during the training process. First, we train the discriminator with a learning rate of $2\times 10^{-4}$ and the generator with a greater learning rate of $5\times 10^{-4}$. Second, we skip the training of the discriminator when the classification accuracy in the previous iteration exceeds $80\%$. 

\subsection{Testing Details}
Denote the random input vector as $z\in \mathbb{R}^{200}$, LFG as $\mathcal{L}: \mathbb{R}^{200}\mapsto\mathbb{R}^{64\times 64\times 64}$, low pass filter as $\sigma: \mathbb{R}^{64\times 64\times 64}\mapsto \mathbb{R}^{64\times 64\times 64}$, HFG as $\mathcal{H}: \mathbb{R}^{64\times 64\times 64}\mapsto \mathbb{R}^{64\times 64\times 64}$, generated SDF as $\mathcal{S}\in\mathbb{R}^{64\times 64\times 64}$, we describe the conjoined test process for shape generation as:
\begin{align}
\mathcal{S} = \underbrace{\sigma(\mathcal{L}(z))}_{\text{Low-frequency Part by LFG}}+\underbrace{\mathcal{H}(\sigma(\mathcal{L}(z)))}_{\text{High-frequency Part by HFG}}
\end{align}

A simple low-pass filter, as implemented in this example, can be achieved by:
\begin{align}
\sigma(:) = \mathcal{F}^{-1}(\mathcal{P}_{\omega}(\mathcal{F}(:)))
\end{align}
Where $\mathcal{F}$ and $\mathcal{F}^{-1}$ denote the forward and inverse Fourier Transform, $\mathcal{P}_f$ denotes the process of padding all frequency modes greater than $\omega$ to zero. As a practical note, for better visual appearance, symmetry of geometric shapes of interest can be exploited and enforced during test time. Results in this study are produced by enforcing mirror symmetry at test time.

\section{Results}
\label{sec:result}
In this section we present the results from this study. We first give a qualitative display of generated results and compare them with the previous state-of-the-art. We also display a rendered result of the resulting mesh after going through an automated mesh processing pipeline to illustrate the quality of the results and its potential in CG. We then show the detail enhancing effects of our high-frequency generator to illustrate its effectiveness. Last but not least, we show that our network possess the same latent space properties that allow shape semantic arithmetics and shape interpolations.
\subsection{Shape Generation}
A different model is trained for each individual class in the ShapeNet dataset. In this study, we trained for three object classes namely the car, chair and table classes in ShapeNet. A sample from each classes is displayed in Figure \ref{fig:collage}. A random sample is reproduced from \cite{3dgan} and displayed for comparison. Results generated from this study is qualitatively more visually appeasing, partially due to the smooth meshed surfaces extracted from signed distance function fields, but also due to detail enhancing from HFG that creates sharp high frequency local features.

\begin{figure}[t!]
\centering
\includegraphics[width=.8\textwidth]{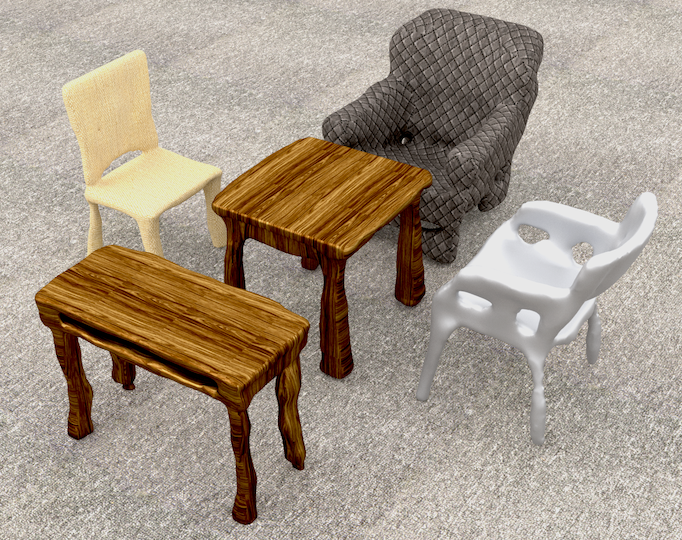}
\caption{A rendering of some examples generated in the chair and table classes. The mesh are post-processed by conventional CG techniques, namely marching cubes algorithm (surface mesh extraction), Laplace smoothing, uv-unwrap and texture mapping. Scene is rendered using Blender \cite{blender}}.
\label{fig:furniture}
\end{figure}

A rendered image of chair and table models created by this learning framework is given in Figure \ref{fig:furniture}. This rendering is to illustrate the quality and refined detail levels of the models generated, and it's potential for real-world applications.

\subsection{Detail Enhancement}
\begin{figure}[t!]
\centering
\includegraphics[width=.32\textwidth]{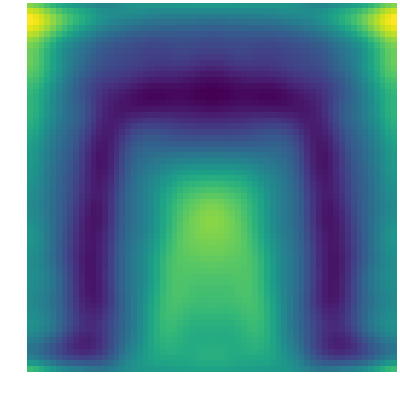}
\includegraphics[width=.32\textwidth]{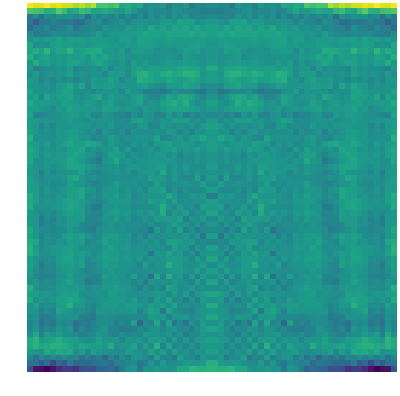}
\includegraphics[width=.32\textwidth]{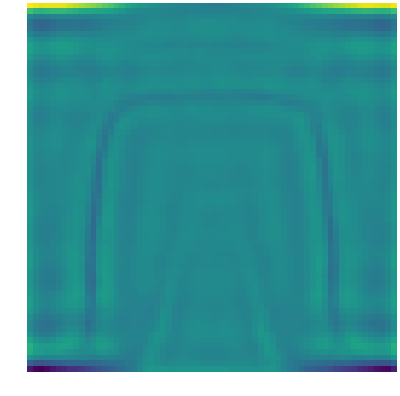}
\caption{Left: Low frequency portion (retaining modes $[-8, 8]$) of LFG output sample. Center: high frequency portion (retaining modes $[-32, -9]\cup[9, 31]$) of LFG output sample. Right: high frequency generated by HFG from leftmost as input. Distance functions are generated at $64^3$ resolution level.}
\label{fig:lfhf}
\end{figure}

\begin{figure}[t!]
\centering
\includegraphics[width=.32\textwidth]{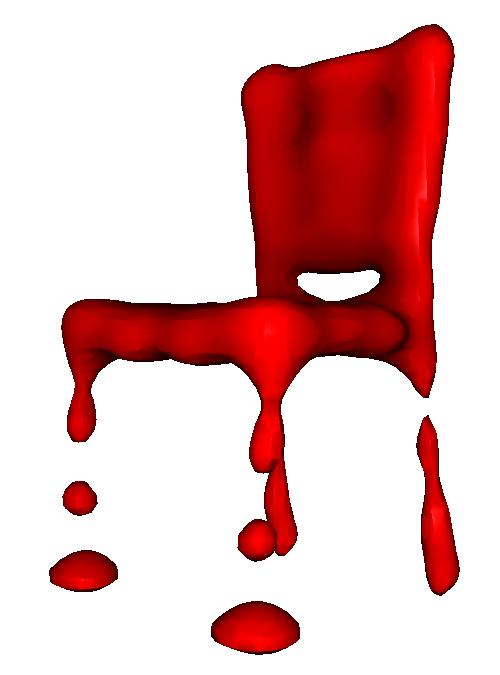}
\includegraphics[width=.32\textwidth]{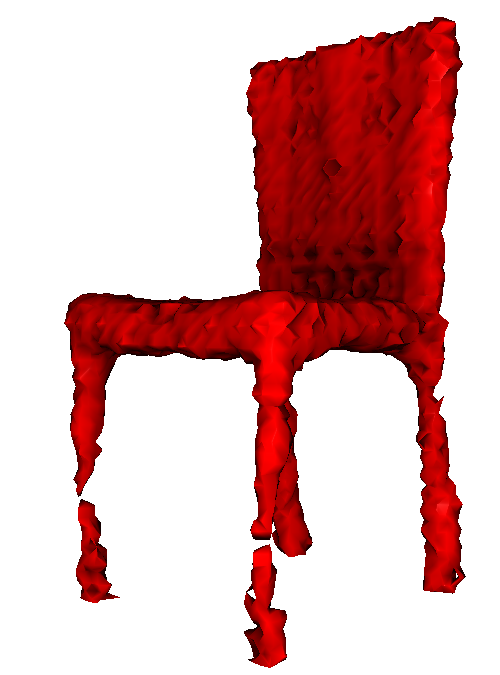}
\includegraphics[width=.32\textwidth]{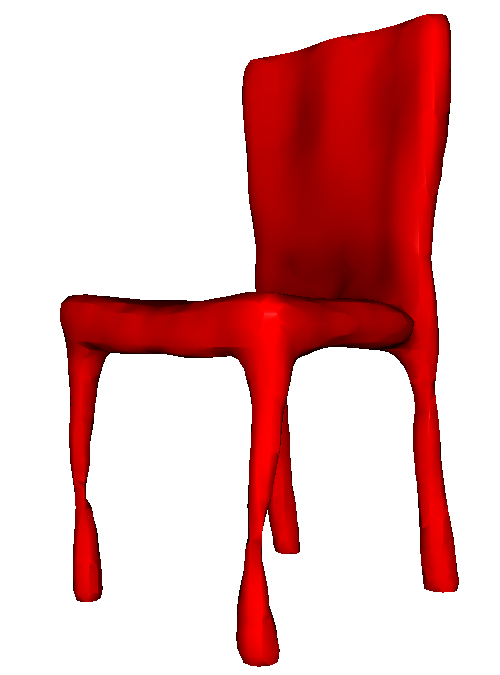}
\caption{Left: surface extracted from Low frequency portion (retaining modes $[-8, 8]$) of LFG output sample. Center: original low and high frequency combined LFG output sample. Right: final output consisting of LFG low frequencies plus HFG high frequency output.}
\label{fig:lfhg_surf}
\end{figure}

The effects of detail enhancement is illustrated in Figure \ref{fig:lfhf}. The figure shows a representative sample generated for the chair class. The LFG network closely mimics the network architecture of \cite{3dgan}. By separating out the low and high frequencies of the output, it is clear that the generator is able to learn low frequency functions quite well, but fail to learn meaningful high frequency functions. Though the GAN and up-convolutional network is appealing for it's one-step generation of 3D shape functions (combined high and low frequencies), the higher frequencies are noisy and not meaningful. Hence the true resolution of the output (modes of meaningful frequencies) via the network architecture in LFG and \cite{3dgan} is lower than the physical resolution (number of voxels). The HFG is successful in mapping lower frequencies to it's corresponding higher frequencies to achieve super-resolution in frequency domain. Figure \ref{fig:lfhg_surf} illustrates the noise in the original output from LFG, effects of low-pass filtering, and HFG's super-resolution effects for generating sharp features.

\subsection{Shape Arithmetics}
Semantic shape arithmetics can be performed for latent shape vectors. Linear interpolations of latent space vectors can result in morphed output shapes.

\begin{figure}[t]
\centering
\includegraphics[width=.1\textwidth]{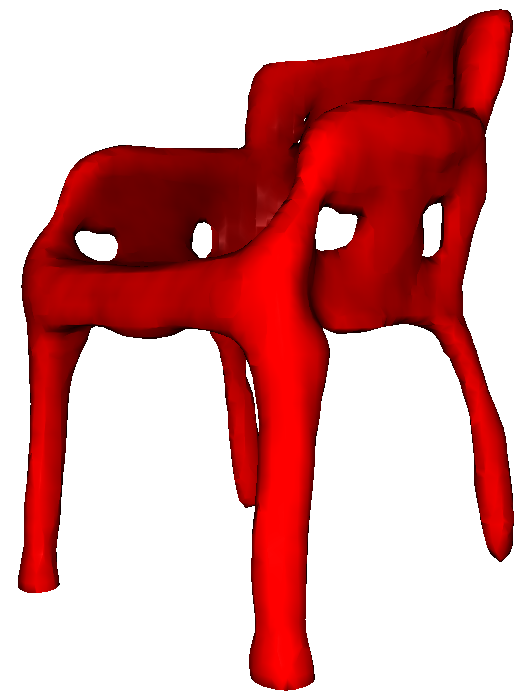}
\includegraphics[width=.1\textwidth]{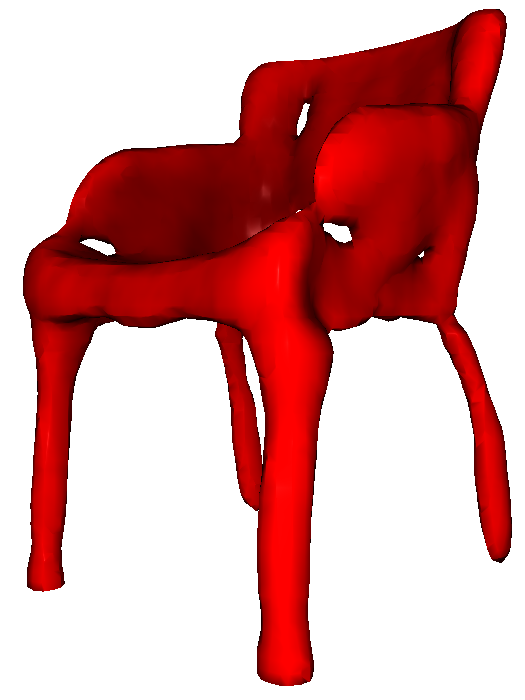}
\includegraphics[width=.1\textwidth]{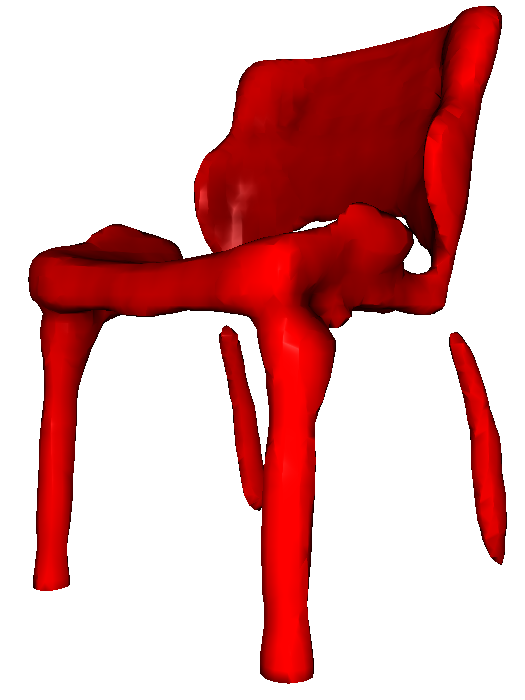}
\includegraphics[width=.1\textwidth]{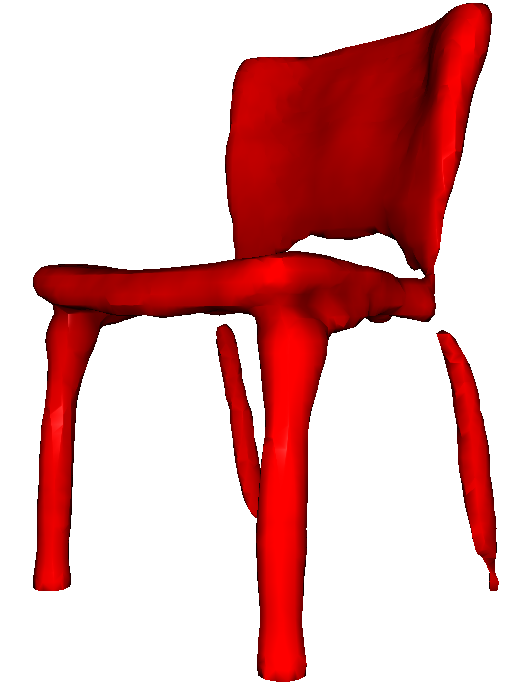}
\includegraphics[width=.1\textwidth]{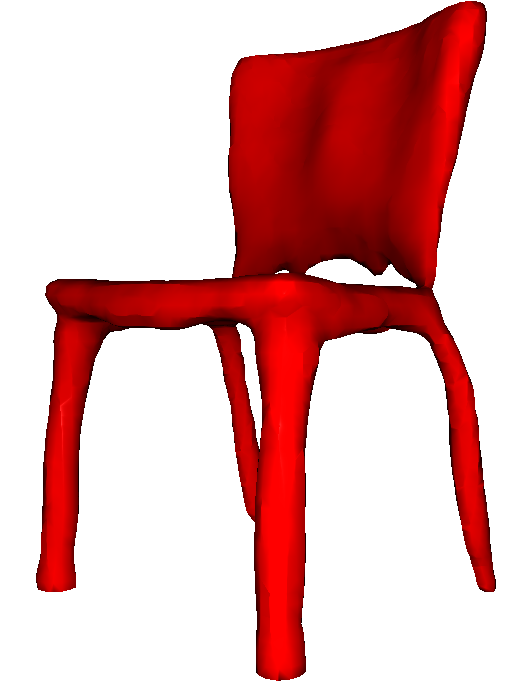}
\includegraphics[width=.1\textwidth]{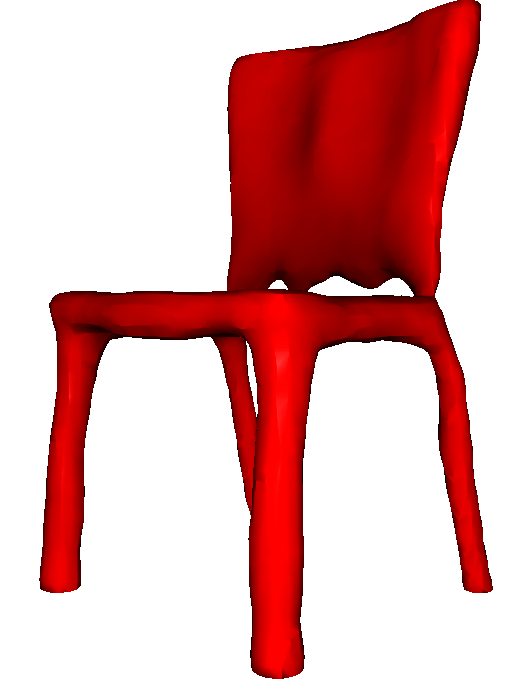}
\includegraphics[width=.1\textwidth]{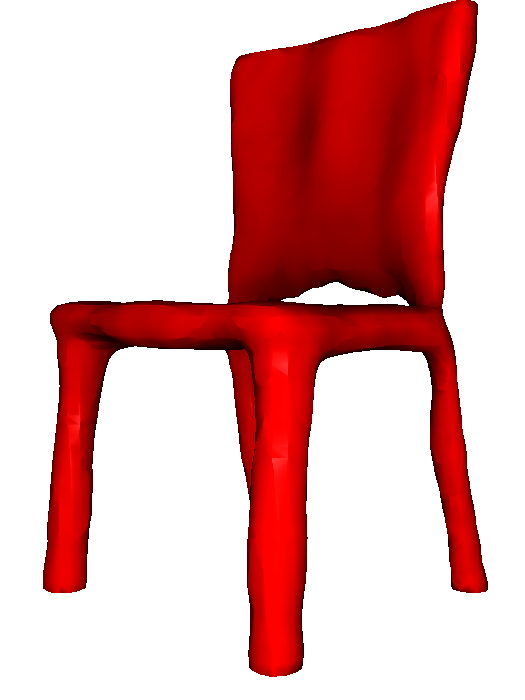}
\includegraphics[width=.1\textwidth]{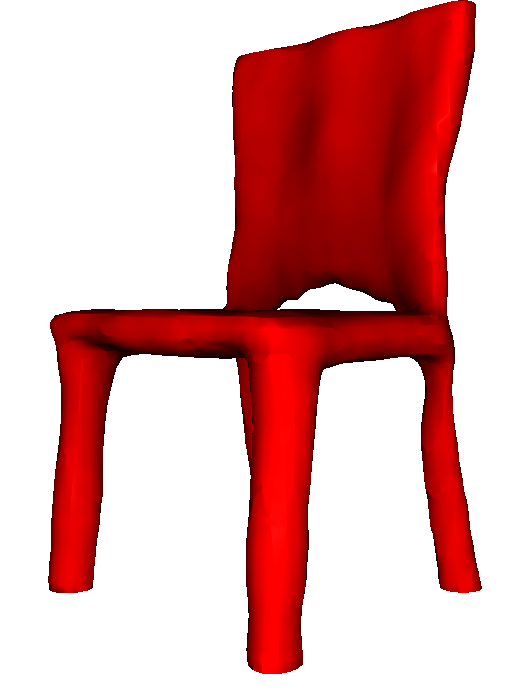}
\includegraphics[width=.1\textwidth]{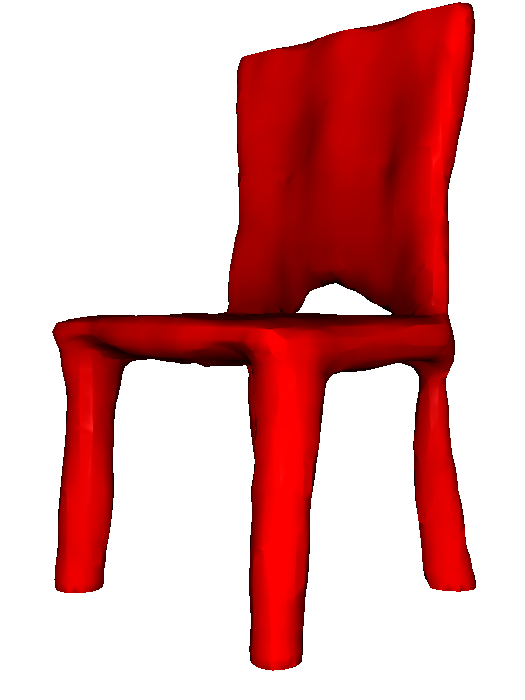}
\caption{Continuous morphing of output shapes achieved by linear interpolation of shape vectors.}
\end{figure}

\section{Conclusion}
\label{sec:conclusion}
We have shown in this study that the signed distance function is a data representation that provides better resolved results for shape generation using 3D-GAN, compared to previous methods that utilize binary voxel representations. We have shown that generated shapes have smooth surface properties, as well as refined details. They are conceptually novel and different from examples given in the training set. We have further shown that manipulations in latent vector space can result in semantic arithmetic operations in the physical space. Our main contribution, however, is in the use of a hierarchical architecture of two networks, namely the Low-Frequency Generator and High-Frequency Generator for the generation of refined surfaces of high quality.
\subsubsection*{Acknowledgments}
We would like to acknowledge the helpful input from Alexei Efros. We acknowledge help with initial stage exploratory work from Madeleine Traverse. We are also grateful for Microsoft Azure for sponsoring GPU computational resources for initial stages of this project.
\bibliography{ref.bib}
\bibliographystyle{unsrtnat}

\end{document}